\documentclass[a4paper,12pt]{article}
\usepackage{amsmath}
\usepackage{amsfonts}
\usepackage{amssymb}
\usepackage{graphicx}
\usepackage{cite}
\usepackage{wasysym}

\begin{document}

\date{}

\title{\bf  Density dependence of Nuclear Symmetry Energy}

\author{B. Behera \footnote{School of Physics, Sambalpur University, Jyotivihar, Burla, Sambalpur, Odisha-768019, INDIA}, T. R. Routray \footnote{School of Physics, Sambalpur University, Jyotivihar, Burla, Sambalpur, Odisha-768019, INDIA, trr1@rediffmail.com} and S. K. Tripathy\footnote{Department of Physics, Indira Gandhi Institute of Technology, Sarang, Dhenkanal, Odisha-759146, INDIA, tripathy\_ sunil@rediffmail.com}      
}

\maketitle

\begin{abstract} High density behaviour of nuclear symmetry energy is studied on the basis of a stiffest density dependence of asymmetric contribution to energy per nucleon in charge neutral $n+p+e+\mu$ matter under beta equilibrium. The density dependence of nuclear symmetry energy obtained in this way is neither very stiff nor soft at high densities and is found to be in conformity with recent observations of neutron stars 

\end{abstract}

%\textbf{AMS Classification number}: 83F05



\section{Introduction}

The behaviour of nuclear symmetry energy $E_s(\rho)$ at high densities $\rho$ is very crucial for the understanding of many astrophysical phenomena such as chemical composition and related mechanisms in the process of formation of neutron stars. Significant progress has been made to constrain the nuclear symmetry energy at low densities from dynamical behaviour \cite{Tsang09, Tsang12}, resonances and excitations \cite{Piek12, Tami11, Roca13, Roca13a, Zhang14}, static properties of finite nuclei \cite{Gor09, Wang14, Tian14, Mondal16a, Chen15} and neutron skin thickness \cite{Warda09, Wang13, Zhang13, Mondal16, BKA12, BKA13}. However, the high density behaviour of $E_s(\rho)$  is not yet well determined, particularly, at densities much above the saturation point $\rho_0$ and at densities that occur in neutron star core. There is no experimental evidence (from terrestrial laboratories) about the density dependence of $E_s(\rho)$  at these densities. To obtain any description of high density behaviour of $E_s(\rho)$  we are therefore forced to extrapolate a theory which is well tested merely around saturation density $\rho_0$. These extrapolated results of $E_s(\rho)$ at high densities as predicted by different theoretical models are extremely divergent and even contradictory \cite{Dieper03,Liu02, Stone03, Horo01, Lat00 }. However, recent observations of neutron star cooling \cite{Blaske04, Popov06, Kolo05, Page04} suggest that fast cooling through direct URCA (DU) processes should not occur at least in typical neutron stars with masses in the range $1- 1.5 M_{\astrosun}$  which means that $E_s(\rho)$  should not be very stiff at high densities. Arguments in support of non-occurrence of DU processes in neutron stars has also been arrived at in the works of Page et al. \cite{Page04}, Basu et al. \cite{DNB08} and Centelles et al. \cite{Cent09}. Basu et al. have examined the compact star cooling phenomenology using equation of state (EOS) of asymmetric nuclear matter (ANM) obtained from density-dependent M3Y interaction (DDM3Y). Centelles et al. \cite{Cent09} have reached in this conclusion from the calculation of transition density between crust and core in their work on probing the nuclear symmetry energy from the measurements of neutron skin thickness of nuclei across the mass table. In some of our recent works, we have also investigated the temperature and density dependence of nuclear symmetry energy using  finite range effective interactions \cite{SKT09, SKT11}.

In the present work, we study the high density behaviour of nuclear symmetry energy on the basis of a stiffest density dependence of asymmetric contribution to the energy per nucleon in beta equilibrated $n+p+e+\mu$  matter, i.e., neutron star matter (NSM) and examine its relevance to the DU constraint mentioned above.
Our starting point is the energy per nucleon $E (\rho, Y_p)$  in ANM at zero temperature and can be written as, 

\begin{equation}
E (\rho, Y_P)= \frac{1}{\pi ^2 \rho}\left[f(k_n)+f(k_p)\right]+V(\rho, Y_p),
\end{equation}
where, $\rho=\rho_n + \rho_p$  is the nucleon density, $Y_p=\frac{\rho_p}{\rho}$  is the proton fraction and $k_{n,p}$  are the neutron and proton Fermi momenta corresponding to densities $\rho_n$  and $\rho_p$ . The first term involving the square brackets is the kinetic part and $V(\rho, Y_p)$  is the interaction part of $E (\rho, Y_p)$ . The kinetic part is treated in the relativistic Fermi gas model and the functional $f(k_i)$ with $i=n,p$ is defined as \cite{Behera02, Behera05},

\begin{equation}
f(k_i)=\int_0^{k_f} \left(c^2\hbar^2 k^2+M^2c^4\right)^{\frac{1}{2}}k^2 dk,
\end{equation}
In general $V(\rho, Y_p)$  has a complicated dependence on $\rho$  and $Y_p$  due to finite range exchange interactions between nucleons. However, the isospin symmetry allows to decompose it in a series of even integral powers of $(1-2Y_p)$  where the contributions coming from higher order terms are very small compared to the leading quadratic term \cite{Lee98}. But depending on the nature of finite range exchange interactions operating between like and unlike nucleons, the higher order terms can have importance in certain situations such as the onset of DU processes in NSM \cite{Steiner06}. Also the higher order terms may have some relevance in predicting the masses of finite nuclei in addition to other effects [\cite{Wang15} and references therein]. However, to maintain simplicity, we ignore higher order terms so that $E(\rho, Y_p)$  takes the form
\begin{equation}
E(\rho, Y_p)= \frac{1}{\pi^2 \rho}\left[f(k_n)+f(k_p)\right]+V_0(\rho) +(1-2Y_p)^2 V_s(\rho),
\end{equation}
where $V_0(\rho)$ is the interaction energy per nucleon in symmetric nuclear matter (SNM) and $V_s(\rho)$  is the interaction part of nuclear symmetry energy $E_s(\rho)$ . The more important quantity for our purpose is the asymmetric contribution $E_{asy}(\rho, Y_p)$, which can be expressed as 
\begin{eqnarray}
E_{asy}(\rho, Y_p)&=&E(\rho, Y_p)-E_0(\rho)\\\nonumber
&=&\frac{1}{\pi^2 \rho}\left[f(k_n)+f(k_p)-2f(k_f)\right]+(1-2Y_p)^2 V_s(\rho),
\end{eqnarray}
where, $E_0(\rho)$  is the energy per nucleon and $k_f$  is the Fermi momentum in SNM. We now analyze the high density behaviour of the functional $E_{asy}(\rho, Y_p)$  in NSM to constrain the density dependence of interaction part of nuclear symmetry energy $V_s(\rho)$ . The calculation of $E_{asy}(\rho, Y_p)$ in NSM, i.e.$E_{asy}^{NSM}(\rho, Y_p)$, would require the additional conditions of charge neutrality,
\begin{equation}
Y_p=Y_e+Y_{\mu}
\end{equation}
and beta equilibrium

\begin{equation}
\left[\mu_n(\rho, Y_p)-\mu_p(\rho, Y_p) \right]=\mu_e(\rho, Y_e)=\mu_{\mu}(\rho, Y_{\mu}),
\end{equation}
where, $Y_i=\frac{\rho_i}{\rho}$  with $i=e, \mu$  are the electron and muon fractions in NSM and $\mu_i$  with $i=n, p,e,\mu$   are the respective chemical potentials of neutron, proton, electron and muon. In this work both the leptons, $e$ and $\mu$ are described as relativistic ideal Fermi gas. The difference between neutron and proton chemical potentials in eq.(6) can be given by $-\frac{\partial E(\rho, Y_p) }{\partial Y_p}$ . This can be used to express the beta equilibrium condition in the form,

\begin{eqnarray}
\left\{\left[c^2\hbar ^2 k_f^2(2(1-Y_p))^{\frac{2}{3}}+M^2c^4\right]^{\frac{1}{2}}-\left[c^2\hbar ^2 k_f^2(2Y_p)^{\frac{2}{3}}+M^2c^4\right]^{\frac{1}{2}}\right\} \\ \nonumber
+4(1-2Y_p)V_s(\rho)=\left[c^2\hbar ^2 k_f^2(2Y_e)^{\frac{2}{3}}+m_e^2c^4\right]^{\frac{1}{2}}\\ \nonumber
=\left[c^2\hbar ^2 k_f^2(2Y_{\mu})^{\frac{2}{3}}+m_{\mu}^2c^4\right]^{\frac{1}{2}},
\end{eqnarray}
where, $m_e$  and $m_{\mu}$  are the rest masses of $e$  and $\mu$ .  For each nucleon density $\rho$, eq.(7) is to be solved subject to the condition in eq.(5) to calculate the fractions of $n ,p, e$ and $\mu$   in NSM. The equilibrium proton fraction $Y_p(\rho)$   obtained in this way can be used in eq.(4) to calculate the functional $E_{asy}^{NSM}(\rho, Y_p)$ in NSM.

At a given density $\rho$  we now examine the behaviour of equilibrium proton fraction $Y_p$  and the functional $E_{asy}^{NSM}(\rho, Y_p)$  in NSM with increasing representative values of $V_s(\rho)$. It is obvious from eq.(7) that at a given density $\rho$, the equilibrium proton fraction $Y_p$  in NSM would increase steadily with increasing values of $V_s(\rho)$ . This is illustrated in Figure 1(a) at four different high densities, $\rho = 0.4, 0.8, 1.2$ and $1.6 fm^{-3}$. For $n+p+e+\mu$  matter in beta-equilibrium, DU processes become operative when the proton fraction exceeds a critical value calculated from the condition $Y_n^{\frac{1}{3}}=Y_p^{\frac{1}{3}}+Y_e^{\frac{1}{3}}$  and/or $Y_n^{\frac{1}{3}}=Y_p^{\frac{1}{3}}+Y_{\mu}^{\frac{1}{3}}$. It should be mentioned here that the critical value of proton fraction in $n+p+e+\mu$  matter in beta-equilibrium is around $0.11$. This critical value of $Y_p$  for a given density $\rho$  corresponds to a limiting value of $V_s(\rho)$  as can be seen from Fig.1(a). The limiting values of $V_s(\rho)$ beyond which the proton fraction would exceed its critical value to allow DU processes at different high densities $\rho$  are denoted by $[V_s(\rho)]_{threshold}$. The density dependence of  $[V_s(\rho)]_{threshold}$ is given in Fig.1(b) and marked as curve A. Curves of $V_s(\rho)$ which pass above curve A would allow DU processes to occur, whereas, DU processes can not take place for those passing below it.  The behaviour of the functional $E_{asy}^{NSM}(\rho, Y_p)$  in NSM at different given densities $\rho$  with increasing representative values of $V_s(\rho)$  is shown in Figure 1(c). Unlike the behaviour of  $Y_p(\rho)$ given in Fig.1(a),  $E_{asy}^{NSM}(\rho, Y_p)$ shows a much smaller variation over a wide range of increasing values of  $V_s(\rho)$. It is further seen from Fig.1(c) that for a given density $\rho$  the functional  $E_{asy}^{NSM}(\rho, Y_p)$ attains a maximum value corresponding to a critical value of $V_s(\rho)$ , i.e., $[V_s(\rho)]_{critical}$. The density dependence of $[V_s(\rho)]_{critical}$ is given in Fig.1(b) and marked as curve B. In Fig.1(b) we have also shown another curve of $V_s(\rho)$  below curve B, namely $[V_s(\rho)]_{soft}$  and marked as C, such that curves A and C are equidistant to B. The high density behaviour of $E_{asy}^{NSM}(\rho, Y_p)$  calculated with the three different curves of $V_s(\rho)$, namely A, B and C, are given in Figure 1(d). We note that the curve of $[V_s(\rho)]_{critical}$  gives rise to the stiffest high density behaviour of $E_{asy}^{NSM}(\rho, Y_p)$  over the entire range of density shown. The two other curves of $E_{asy}^{NSM}(\rho, Y_p)$  obtained with  $[V_s(\rho)]_{threshold}$ and $[V_s(\rho)]_{soft}$  are identical and very close to the stiffest one mentioned above. Thus all curves of $V_s(\rho)$  which pass in between A and C in Fig.1(b) would give almost the same high density behaviour of $E_{asy}^{NSM}(\rho, Y_p)$. On the other hand, when we calculate $E_{asy}^{NSM}(\rho, Y_p)$  with curves of $V_s(\rho)$  on the higher and higher side of curve A / lower and lower side of curve C, its high density behaviour will deviate more and more from the stiffest one. This is also shown in Fig.1(d) by considering two equidistant curves of $V_s(\rho)$, one above A and another below C. Comparing Figs.1(b) with 1(d) we note that under the condition of beta-equilibrium, the range of the functional $E_{asy}^{NSM}(\rho, Y_p)$  shows a much smaller variation over a wide range of density than exhibited by the corresponding curves of $V_s(\rho)$ . In view of this, the high density behaviour of the functional  $E_{asy}^{NSM}(\rho, Y_p)$ can be regarded as universal to a good approximation in case of nuclear EOSs for which $V_s(\rho)$  is neither very stiff nor soft compared to the curve of $[V_s(\rho)]_{critical}$. Nuclear EOSs for which the density dependence of $V_s(\rho)$   pass in between curves A and C in Fig.1(b) do not allow DU processes to occur and simultaneously fulfil  the universal high density behaviour of $E_{asy}^{NSM}(\rho, Y_p)$ . Klahn et. al. in their work \cite{Klahn06} on formulating a scheme for testing nuclear matter EOSs at high densities using constraints from neutron star phenomenology and flow data from heavy-ion collisions have noticed this universal high density behaviour of the functional  $E_{asy}^{NSM}(\rho, Y_p)$ for five numbers of EOSs out of the total eight relativistic EOSs considered in their work. All the curves of $E_{asy}^{NSM}(\rho, Y_p)$  for these five EOSs form a narrow band over a wide range of density and these EOSs are also found to satisfy the DU constraint obtained from present cooling phenomenology \cite{Blaske04, Popov06, Kolo05, Page04}. The curves for rest three EOSs deviates widely from this universal behaviour and also do not satisfy the DU constraint. This universal high density behaviour of $E_{asy}^{NSM}(\rho, Y_p)$  has been used in a recent work \cite{Behera07} in constraining the parameters of the finite range interaction used in the study of EOS of ANM and recent observations on neutron star bulk properties. The EOS of ANM constrained by using this universal high density behaviour of $E_{asy}^{NSM}(\rho, Y_p)$  conforms to the DU constraint alongwith providing a satisfactory description of the observed neutron star mass, radius and, in particular, the free baryonic ~ gravitational mass relationship of pulsar B in the double pulsar system J0737-3039 \cite{Behera02, Behera05}. It is to note here that the values of the functional $E_{asy}^{NSM}(\rho, Y_p)$  corresponding to the universal behaviour obtained as a function of density in the cases of the EOSs constructed from relativistic and non-relativistic considerations in Refs.\cite{Behera02} and \cite{Behera05}, respectively, are in close agreement with the model independent values found in this work. In this context it can be mentioned here that although recent observations of neutron star cooling \cite{Blaske04, Popov06, Kolo05, Page04, DNB08, Cent09} suggest that fast cooling through DU processes should not occur in neutron stars and  the                                                                                                                                    universal high density behaviour of the functional $E_{asy}^{NSM}(\rho, Y_p)$  leads to a density dependence of $V_s(\rho)$  in consistent with these observations. In the remaining part of this work we use the stiffest high density behaviour of $E_{asy}^{NSM}(\rho, Y_p)$  to constrain the density dependence of interaction part of nuclear symmetry energy $V_s(\rho)$.

\begin{figure}
\begin{center}
\includegraphics[width=1\textwidth]{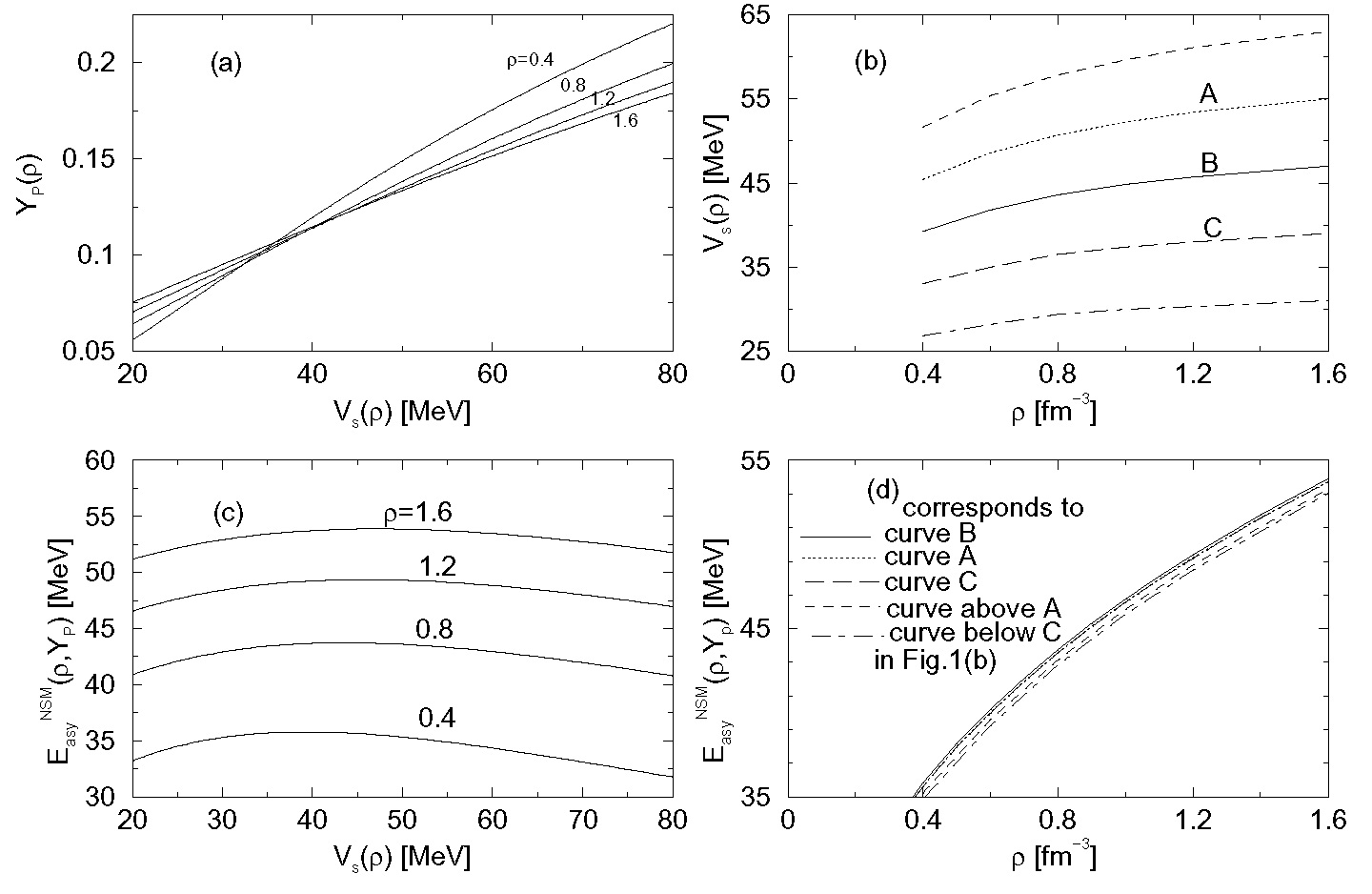}
\caption{(a) Steady increase of equilibrium proton fraction $Y_p(\rho)$  with increasing values of $V_s(\rho)$ is shown at four different high densities, $\rho = 0.4, 0.8, 1.2 $ and $1.6 fm^{-3}$.  (b) High density behaviour of the functionals $[V_s(\rho)]_{threshold}$, $[V_s(\rho)]_{critical}$  and $[V_s(\rho)]_{soft}$  marked as curves A, B and C, respectively, along with two equidistant curves of $V_s(\rho)$, one above A and other below C (see the text for details).  (c) Variation of the functional $E_{asy}^{NSM}(\rho, Y_p)$  in NSM with increasing values of $V_s(\rho)$ at four different high densities,  $\rho = 0.4, 0.8, 1.2 $ and $1.6 fm^{-3}$ and   (d) High density behaviour of  $E_{asy}^{NSM}(\rho, Y_p)$ calculated with the five different curves of $V_s(\rho)$  in Fig.1(b). }
\end{center}
\end{figure}

Up to this stage the high density behaviour of $V_s(\rho)$ as well as the approximate universal density dependence of $E_{asy}^{NSM}(\rho, Y_p)$  have been described in a model independent way in the sense that no specific functional form of $V_s(\rho)$  involving adjustable parameters was used for the purpose. However, this model independent description of high density behaviour of $[V_s(\rho)]_{critical}$  gives significantly higher values when curve B in Fig.1(b) is extrapolated to the low density region and are not in well agreement with those predicted by nuclear structure as well as reaction studies \cite{Dieper03,Moller95, Pomor03, Khoa05, Khoa07, DVS07}. For example, the value of $[V_s(\rho)]_{critical}$  obtained in this way varies from 32.8 to 33.6 MeV for a value of $\rho_0$  in the range 0.15 to 0.17 $fm^{-3}$ and can be compared with the standard value of around 18 MeV \cite{TM07}. In view of this it is necessary to smoothly connect the low density behaviour of $V_s(\rho)$  obtained from nuclear structure and reaction studies to the high density behaviour of curve B shown in Fig.1(b). For this purpose we need a functional form of $V_s(\rho)$. In this context we note that the curve of  $[V_s(\rho)]_{critical}$  in Fig.1(b) asymptotically approaches a definite value when extrapolated to very large densities.  Any functional form chosen for $V_s(\rho)$  must therefore fulfill this feature in the high density region. On the other hand in the low density region, i.e.,$\rho \rightarrow 0$ , the functional  $V_s(\rho)$ should behave as \cite{DNB08, Chen05}
\begin{equation}
V_s(\rho) \xrightarrow{\rho \rightarrow 0}  b \rho,
\end{equation}
where,  $b$ is a constant. Out of the several functional forms of $V_s(\rho)$  those can be constructed consistent with the behaviour in both the asymptotic regions discussed above, we consider the following two representative forms,

\begin{eqnarray}
V_s(\rho)&=& A\left(\frac{\rho}{1+a\rho}\right),\\
V_s(\rho)&=& D \left[1-exp(-d\rho)\right],
\end{eqnarray}
and examine their ability to reproduce the experimentally extracted constraints on nuclear symmetry energy. The two adjustable parameters involved in each of these two functional forms of $V_s(\rho)$  can be described by $V_s(\rho_0)$  and  $V'_s(\rho_0)=\rho \frac{dV_s(\rho)}{d \rho}\mid_{\rho=\rho_0} $ which are directly related to the symmetry energy  $E_s(\rho_0)$  and the slope parameter $L(\rho_0)=3 \rho\frac{dE_s(\rho)}{d\rho}\mid _{\rho=\rho_0}$  at saturation density . In this context we note that while different theoretical models give similar results of $V_s(\rho_0)$  they differ widely in the values of $V'_s(\rho_0)$ \cite{Stone03, Brown00}. In view of this we consider the standard range of $V_s(\rho_0)$  in between 16 to 20 MeV and vary $V'_s(\rho_0)$  to find out the critical values corresponding to the stiffest behaviour of the functional $E_{asy}^{NSM}(\rho, Y_p)$  over a wide range of high densities. The critical values of $V'_s(\rho_0)$  obtained in this way for both the functional forms of $V_s(\rho)$  in eqs.(9) and (10) are listed in Table 1. In calculating the results given in table 1 we have taken standard values of $Mc^2=939 MeV$, energy per nucleon in SNM $E_0(\rho_0)=923 MeV$ and $\left(c^2\hbar^2 k_f^2+M^2c^4\right)^{\frac{1}{2}} = 976 MeV$ (corresponding to $\rho_0= 0.1658 fm^{-3}$). The density dependence of $V_s(\rho)$  for different sets in table 1 are shown in Figure 2(a). All the four curves cross over at $\rho= 1.15 fm^{-3}$  which is a consequence of the stiffest high density behaviour of the functional $E_{asy}^{NSM}(\rho, Y_p)$ calculated with these different $V_s(\rho)$. As a result of this, the stiffness of the curves for $V_s(\rho)$ increases with increase in the standard value of $V_s(\rho_0)$ for each of the two functional forms in eqs.(9) and (10) in the region $\rho < 1.15 fm^{-3} $  and conversely for $\rho > 1.15 fm^{-3} $. In this context it may be noted that the stiffness of the curve of $V_s(\rho)$ are very often characterized by three parameters, namely, the symmetry energy $E_s(\rho_0)$, the slope parameter  $L(\rho_0)$ and isospin dependent part of isobaric incompressibility $K_{asy}(\rho_0)=\left[9\rho^2 \frac{d^2E_s(\rho)}{d\rho^2}-18\rho\frac{dE_s(\rho)}{d\rho}\right]_{\rho=\rho_0}$  at saturation density. The stiffest high density behaviour of the functional  $E_{asy}^{NSM}(\rho, Y_p)$ for all sets of $V_s(\rho_0)$ in table 1 constrains the parameters $L(\rho_0)$ and $K_{asy}(\rho_0)$  in the ranges  $62 MeV\leq L(\rho_0)\leq  70 MeV$ and $- 522 MeV\leq  K_{asy}(\rho_0) \leq- 449 MeV$. These results are well within the ranges extracted from a combination
of analysis of neutron skin thickness in finite nuclei and isospin diffusion in heavy ion collision \cite{Cent09,DVS07, TM07, Chen05, Typel01, Chen05a, Li06}. It is worth to mention here that, Agrawal et al. constrained this parameter from neutron skin of heavy nuclei to be $L_0=59\pm 13.0 MeV$ \cite{BKA12, BKA13}. Analysis from isovector giant dipole and quadrupole resonances of $^{208}Pb$ nucleus provide $L_0=43\pm 26.0 MeV$ and $L_0=37\pm 18.0 MeV$ respectively \cite{Roca13, Roca13a}. Zhang and Chen from their studies on electric dipole polarisability in $^{208}Pb$ constrained $L_0$ in the range $L_0=47.3\pm 7.8 MeV$ \cite{Zhang14}. From the values of $L(\rho_0)$ and $K_{asy}(\rho_0)$  in table 1 as well as from Fig.2(a) it can be seen that the exponential form with $V_s(\rho_0)=20 MeV$ (Set-IIB) and the other form with $V_s(\rho_0)=16 MeV$ ( Set-IA) constitute the two extreme curves for $V_s(\rho)$. The density dependence of nuclear symmetry energy $E_s(\rho)$  obtained with these two extreme curves of  $V_s(\rho)$ are given in Figure 2(b). The high density behaviour of $[E_s(\rho)]_{threshold}$  as well as $[E_s(\rho)]_{critical}$ obtained from the two curves A and B in Fig.1(b) are also shown in Fig.2(b) for comparison. The curve of $E_s(\rho)$  obtained with Set-IIB is in good agreement with the curve of $[E_s(\rho)]_{critical}$  over a wide range of high densities. Moreover, the two curves of $E_s(\rho)$ obtained from Sets-IA and IIB are significantly below the curve of $[E_s(\rho)]_{threshold}$. In view of this none of the four different sets of $V_s(\rho)$  listed in table 1 allows DU processes to occur. The density dependence of the functional $E_{asy}^{NSM}(\rho, Y_p)$  calculated with the two extreme curves of $V_s(\rho)$, namely, set IA and IIB are given in Figure 2(c). The stiffest high density behaviour of $E_{asy}^{NSM}(\rho, Y_p)$  obtained with $V_s(\rho)$  and given in Fig.1(b) is also shown in Fig.2(c) for comparison. The three curves of  $E_{asy}^{NSM}(\rho, Y_p)$ are almost the same over a wide range of high densities.

\begin{table}
\caption{Critical values of $V'_s(\rho_0)$  for the stiffest behaviour of the functional $E_{asy}^{NSM}(\rho, Y_p)$  over a wide range of high densities obtained with different standard values of $V_s(\rho_0)$  for both the parametrized versions of $V_s(\rho)$  given in eqs.(9) and (10). The corresponding values of nuclear symmetry energy $E_s(\rho_0)$, slope parameter $L(\rho_0)$ and isospin part of isobaric incompressibility $K_{asy}(\rho_0)$  at saturation density are also listed.}
\centering
\begin{tabular}{l|c|c|c|c}
\hline \hline
			&	\multicolumn{2}{c}{ $V_s (\rho)$ in eq. (9)} 		& 	\multicolumn{2}{c}{ $V_s (\rho)$ in eq. (10)}			\\
	\hline
	[MeV]			&	Set- IA		&	Set- IB		&	Set- IIA		   & Set- IIB	\\
\hline
$V_s(\rho_0)$ 		&    16	  	&	20 		& 16				&	20	\\
$V'_s(\rho_0)$		&	12.2	&	13.2	& 13.1				&	15.0	\\
$E_s(\rho_0)$		&	28.8	&	32.8	& 28.8				&	32.8	\\
$L(\rho_0)$ 		&	61.5	&	64.5	& 64.2				&	69.9	\\
$K_{asy}(\rho_0)$	&	-449	&	-496 	& -459				&	-522	\\
\hline
\end{tabular}
\end{table}

\begin{figure}[h!]
\begin{center}
\includegraphics[width=1\textwidth]{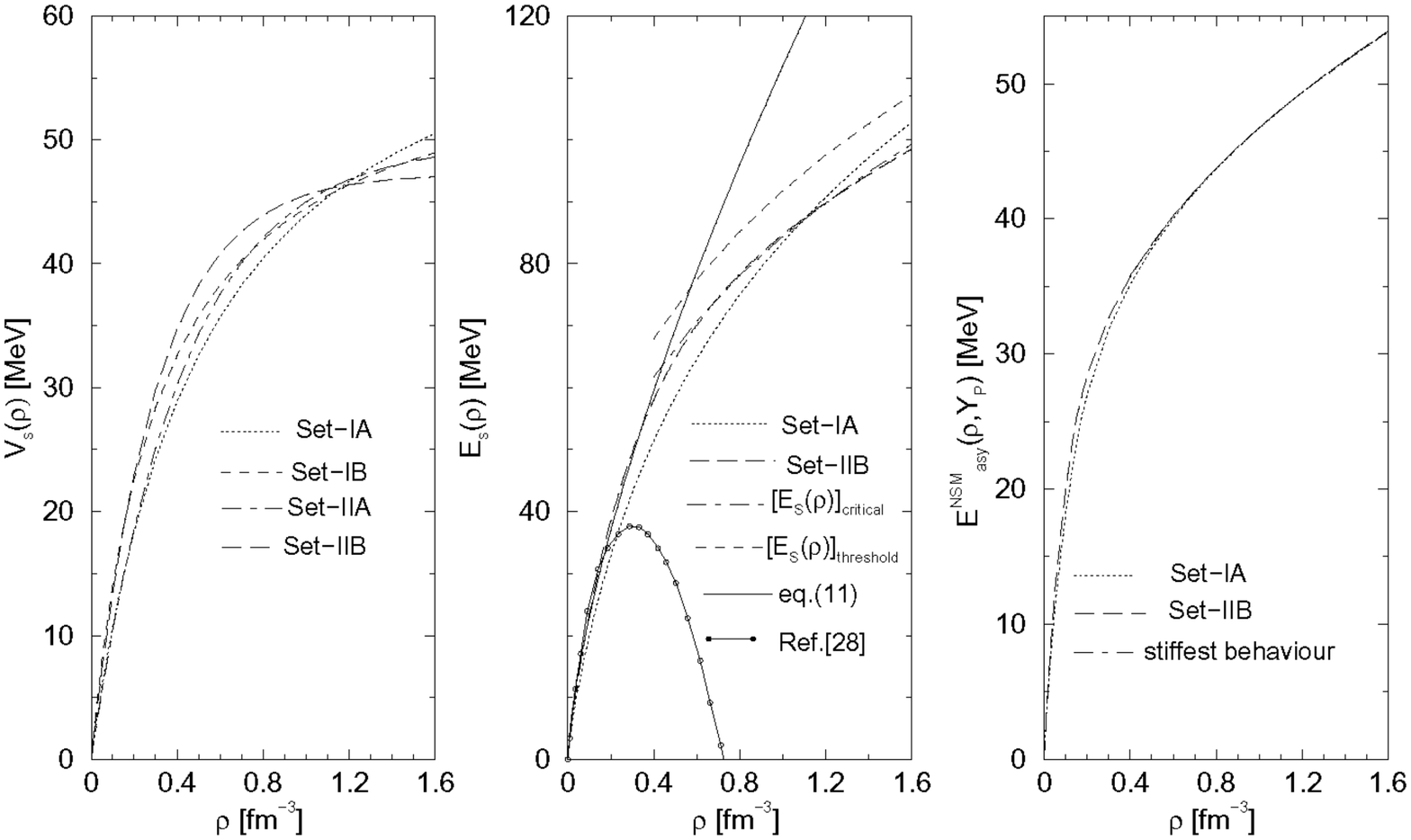}
\caption{(a) Density dependence of $V_s(\rho)$ , calculated with the critical values of $V'_s(\rho)$ for different sets in   table 1, (b)  Density dependence of nuclear symmetry energy  $E_s(\rho)$ obtained from the two extreme curves of $V_s(\rho)$ in Fig.1(b), i.e., Set-IA and IIB. High density behaviour of  $[E_s(\rho)]_{critical}$ and  $[E_s(\rho)]_{threshold}$ (see text for details) as well as the simple result of $E_s(\rho)$  in eq.(11) and that of Ref.\cite{DNB08} are also shown for comparison and (c) Density dependence of the functional $E_{asy}^{NSM}(\rho, Y_p)$  calculated with the two extreme curves of  $V_s(\rho)$ in Fig.2(a) and compared with its stiffest high density behaviour obtained with $[V_s(\rho)]_{critical}$  (see text for details).}
\end{center}
\end{figure}
In context to the high density behaviour of nuclear symmetry energy $E_s(\rho)$  discussed in the present work, it is interesting to note that recent studies on isoscaling of fragment yields \cite{DVS07} in heavy ion collisions have suggested a simple form of  $E_s(\rho)$ at low densities,
%*******************************************************
\begin{equation}
E_s(\rho)=31.6\left(\frac{\rho}{\rho_0}\right)^{0.69}
\end{equation}
%*******************************************************
with $\rho_0=0.16 fm^{-3}$, consistent with the experimental data. This form of $E_s(\rho)$ at low densities is also consistent with the findings of Khoa et. al. \cite{Khoa07} from the study of experimental heavy-ion cross-sections in a charge-exchange reaction with the Hartree-Fock calculation using CDM3Y6 interaction. The density dependence of $E_s(\rho)$  in eq.(11) is also shown in Fig.2(b) for comparison. It is seen that the curve of $E_s(\rho)$  obtained with Set-IIB compares very well with the results in eq.(11) upto a density $\rho \leq 0.4 fm^{-3}$. However, the simple form of  $E_s(\rho)$ in eq.(11) predicts a very stiff behaviour at higher densities which allows DU-processes to occur and does not fulfil the universal high density behaviour of the functional $E_{asy}^{NSM}(\rho, Y_p)$. In order to complete the  discussion on the different types of density dependence of nuclear symmetry energy found in the literature, we also show the symmetry energy curve of Ref. \cite{DNB08} in Fig.2(b). The agreement of our curve for Set-IIB with that of Ref.\cite{DNB08} is quite well upto the normal nuclear matter density $\rho_0$. The symmetry energy curve for the case of Ref.\cite{DNB08} becomes maximum at about  $\rho \approx 0.29 fm^{-3}$ having value $E_{s}(\rho) \approx 37 MeV$  and thereafter it starts decreasing becoming zero at $\rho \approx 0.73 fm^{-3}$, as well as, the DU processes do not occur in this case at any density. The critical density at which the symmetry energy vanishes predicts that the NSM converts to Pure Neutron Matter (PNM) from this density onwards predicting the presence of PNM in the core of the neutron stars for which density in the central region exceeds the value of this critical density. 

In conclusion it can be stated that the consensus view is that nuclear symmetry energy should not be stiff in the high density region. However, currently available experimentally extracted data from heavy-ion collision experiments and informations on composition and cooling mechanism of neutron stars do not provide sufficient ground to pin down the degree of softness of the high density behaviour of nuclear symmetry energy urging upon for more detailed studies on these crucial aspects. In these efforts, it may be suggested to examine the validity of the naturally occurring universality condition, corresponding to the stiffest high density behaviour of $E_{asy}^{NSM}(\rho, Y_p)$ , discussed in the work that restricts the nuclear symmetry energy with in a narrow region.

\end{document}